\documentclass[prb,aps,twocolumn,showpacs,floats]{revtex4}
\usepackage{epsfig,bm,epsf,graphics}
\begin{document}
\draft
\title{Nature of phase transition in magnetic thin films}
\author{X. T. Pham Phu$^{a}$, V. Thanh Ngo$^{b,c}$ and H. T. Diep$^{a}$\footnote{ Corresponding author, E-mail:diep@u-cergy.fr }}
\address{$^{a}$ Laboratoire de Physique Th\'eorique et Mod\'elisation,
CNRS-Universit\'e de Cergy-Pontoise, UMR 8089\\
2, Avenue Adolphe Chauvin, 95302 Cergy-Pontoise Cedex, France\\
$^b$ Institute of Physics, P.O. Box 429,   Bo Ho, Hanoi 10000,
Vietnam\\
$^c$ Asia Pacific Center for Theoretical Physics, Hogil Kim
Memorial Building 5th floor, POSTECH, Hyoja-dong, Namgu, Pohang
790-784, Korea}

\begin{abstract}
We study  the critical behavior of magnetic thin films as a
function of the film thickness.  We use the ferromagnetic Ising
model with the high-resolution multiple histogram Monte Carlo (MC)
simulation. We show that though the 2D behavior remains dominant
at small thicknesses, there is a systematic continuous deviation
of the critical exponents from their 2D values. We observe that in
the same range of varying thickness the deviation of the exponent
$\nu$ is very small from its 2D value, while exponent $\beta$
suffers a larger deviation.  Moreover, as long as the film
thickness is fixed, i. e. no finite size scaling is done in the
$z$ direction perpendicular to the film, the 3D values of the
critical exponents cannot be attained even with very large (but
fixed) thickness. The crossover to 3D universality class cannot
therefore take place without finite size scaling applied in the
$z$ direction, in the limit of numerically accessible thicknesses.
From values of exponent $\alpha$ obtained by MC, we estimate the
effective dimension of the system.   We conclude that with regard
to the critical behavior, thin films behave as systems with
effective dimension between 2 and 3.
\end{abstract}

 \maketitle

\section{Introduction}

During the last 30 years.  physics of surfaces and objects of
nanometric size have attracted an immense interest.  This is due
to important applications in
industry.\cite{zangwill,bland-heinrich}
 An example is the so-called giant
magneto-resistance (GMR) used in data storage devices, magnetic
sensors, ... \cite{Baibich,Grunberg,Fert,review}  In parallel to
these experimental developments, much theoretical
effort\cite{Binder-surf,Diehl} has also been devoted to the search
of physical mechanisms lying behind new properties found in
nanoscale objects such as ultrathin films, ultrafine particles,
quantum dots, spintronic devices etc. This effort aimed not only
at providing explanations for experimental observations but also
at predicting new effects for future experiments.

The physics of two-dimensional (2D) systems is very exciting. Some
of those 2D systems can be exactly solved: one famous example is
the Ising model on the square lattice which has been solved by
Onsager.\cite{Onsager}  This model shows a phase transition at a
finite temperature $T_c$ given by $\sinh^2(2J / k_BT_c)=1$ where
$J$ is the nearest-neighbor (NN) interaction. Another interesting
result is the absence of long-range ordering at finite
temperatures for the continuous spin models (XY and Heisenberg
models) in 2D.\cite{Mermin} In general, three-dimensional (3D)
systems for any spin models cannot be unfortunately solved.
However, several methods in the theory of phase transitions and
critical phenomena can be used to calculate the critical behaviors
of these systems.\cite{Zinn}

This paper deals with systems between 2D and 3D. Many theoretical
studies have been devoted to thermodynamic properties of thin
films, magnetic multilayers,...
\cite{Binder-surf,Diehl,ngo2004trilayer,Diep1989sl,diep91-af-films}
In spite of this, several points are still not yet understood.  We
study here the critical behavior of thin films with a finite
thickness. It is known a long time ago that the presence of a
surface in magnetic materials can give rise to surface spin-waves
which are localized in the vicinity of the surface.\cite{diep79}
These localized modes may be acoustic with a low-lying energy or
optical with a high energy, in the spin-wave spectrum.  Low-lying
energy modes contribute to reduce in general surface magnetization
at finite temperatures. One of the consequences is the surface
disordering which may occur at a temperature lower than that for
interior magnetization.\cite{diep81}  The existence of low-lying
surface modes depends on the lattice structure, the surface
orientation, the surface parameters, surface conditions
(impurities, roughness, ...) etc. There are two  interesting
cases: in the first case a surface transition occurs at a
temperature distinct from that of the interior spins and in the
second case the surface transition coincides with the interior
one, i. e. existence of a single transition. Theory of critical
phenomena at surfaces\cite{Binder-surf,Diehl} and Monte Carlo (MC)
simulations\cite{Landau1,Landau2} of critical behavior of the
surface-layer magnetization at the extraordinary transition in the
three-dimensional Ising model have been carried out.  These works
suggested several scenarios in which the nature of the surface
transition and the transition in thin films depends on many
factors in particular on the symmetry of the Hamiltonian and on
surface parameters.

The aim of this paper is  to investigate the effect of the film
thickness on the critical behavior of the system.  We would like
to see in particular how the  thickness affects the values of
critical exponents.
 To carry out these
purposes, we shall use MC simulations with highly accurate
multiple histogram technique.\cite{Ferrenberg1,Ferrenberg2,Bunker}
We consider here the case of a simple cubic film with Ising model.
For our purpose, we suppose all interactions are the same even
that at the surface.
 This case is the simplest case where there is no surface-localized spin-wave
 modes and there is only a single phase transition at a temperature for the
 whole system (no separate surface phase transition).\cite{diep79,diep81}
  Other complicated cases will
 be left for future investigations.  However, some preliminary discussions
 on this point for complicated surfaces
  have been reported
 in some of our previous papers.\cite{ngo2007,ngo2007fcc}

The paper is organized as follows. Section II is devoted to a
description of the model and method.   Results are shown and
discussed in section III. Concluding remarks are given in section
IV.

\section{Model and Method}
\subsection{Model} Let us consider the Ising spin model on a film
made from a ferromagnetic simple cubic lattice. The size of the
film is $L\times L\times N_z$.  We apply the periodic boundary
conditions (PBC) in the  $xy$ planes to simulate an infinite $xy$
dimension. The $z$ direction is limited by the film thickness
$N_z$.
  If $N_z=1$ then one has a 2D square lattice.

The Hamiltonian is given by
\begin{equation}
\mathcal H=-\sum_{\left<i,j\right>}J_{i,j}\sigma_i\cdot\sigma_j
\label{eqn:hamil1}
\end{equation}
where $\sigma_i$ is the Ising spin of magnitude 1 occupying the
lattice site $i$, $\sum_{\left<i,j\right>}$ indicates the sum over
the NN spin pairs  $\sigma_i$ and $\sigma_j$.

In the following, the interaction between two NN surface spins is
denoted by $J_s$, while all other interactions are supposed to be
ferromagnetic and all equal to $J=1$ for simplicity. Let us note
in passing that in the semi-infinite crystal  the surface phase
transition occurs at the bulk transition temperature when $J_s
\simeq 1.52 J$. This point is called "extraordinary phase
transition" which is characterized by some particular critical
exponents.\cite{Landau1,Landau2}   In the case of thin films, i.
e. $N_z$ is finite, it has been theoretically shown that when
$J_s=1$ the bulk behavior is observed when the thickness becomes
larger than a few dozens of atomic layers:\cite{diep79} surface
effects are insignificant on thermodynamic properties such as the
value of the critical temperature, the mean value of magnetization
at a given $T$, ... When $J_s$ is smaller than $J$, surface
magnetization is destroyed at a temperature lower than that for
bulk spins.\cite{diep81} However, it should be stressed that,
except at the so-called "extraordinary phase
transition",\cite{Landau1,Landau2} the surface criticality has not
been studied as a function of the film thickness.

\subsection{Multiple histogram technique}

The multiple histogram technique is known to reproduce with very
high accuracy the critical exponents of second order phase
transitions.\cite{Ferrenberg1,Ferrenberg2,Bunker}

The overall probability distribution\cite{Ferrenberg2} at
temperature $T$ obtained from $n$ independent simulations, each
with $N_j$ configurations, is given by
\begin{equation}
P(E,T)=\frac{\sum_{i=1}^n H_i(E)\exp[E/k_BT]}{\sum_{j=1}^n
N_j\exp[E/k_BT_j-f_j]}, \label{eq:mhp}
\end{equation}
where
\begin{equation}
\exp[f_i]=\sum_{E}P(E,T_i). \label{eq:mhfn}
\end{equation}

The thermal average of a physical quantity $A$ is then calculated
by
\begin{equation}
\langle A(T)\rangle=\sum_E A\,P(E,T)/z(T),
\end{equation}
in which
\begin{equation}
z(T)=\sum_E P(E,T).
\end{equation}

Thermal averages of physical quantities are thus calculated as
continuous functions of $T$, now the results should be valid over
a much wider range of temperature than for any single histogram.
The $xy$ linear sizes  $L=20, 25, 30, ...,80$ have been used in
our simulations. We have tested that all exponents do not change
in the finite size scaling with $L\geq 30$.  So most of results
are shown for $L\geq 30$ except for $\nu$ where the lowest sizes
$L=20,25$ can be used without modifying its value.

In practice, we use first the standard MC simulations to localize
for each size the transition temperatures $T^E_0(L)$ for specific
heat and $T^m_0(L)$ for susceptibility. The equilibrating time is
from 200000 to 400000 MC steps/spin and the averaging time is from
500000 to 1000000 MC steps/spin. Next, we make histograms at $8$
different temperatures $T_j(L)$ around the transition temperatures
$T^{E,m}_0(L)$ with 2 millions MC steps/spin, after discarding 1
millions MC steps/spin for equilibrating. Finally, we make again
histograms at $8$ different temperatures around the new transition
temperatures $T^{E,m}_0(L)$ with $2\times 10^6$ and $4\times 10^6$
MC steps/spin for equilibrating and averaging time, respectively.
Such an iteration procedure gives extremely good results for
systems studied so far.  Errors shown in the following have been
estimated using statistical errors, which are very small thanks to
our multiple histogram procedure, and fitting errors given by
fitting software.

We have calculated the averaged order parameter $\langle M\rangle$
($M$: magnetization of the film), averaged total energy $\langle
E\rangle$, specific heat $C_v$, susceptibility $\chi$, first order
cumulant of the energy $C_U$, and $n^{th}$ order cumulant of the
order parameter $V_n$ for $n=1$ and 2. These quantities are
defined as

\begin{eqnarray}
\langle E\rangle&=&\langle\cal{H}\rangle,\\
C_v&=&\frac{1}{k_BT^2}\left(\langle E^2\rangle-\langle E\rangle^2\right),\\
\chi&=&\frac{1}{k_BT}\left(\langle M^2\rangle-\langle M\rangle^2\right),\\
C_U&=&1-\frac{\langle E^4\rangle}{3\langle E^2\rangle^2},\\
V_n&=&\frac{\partial\ln{M^n}}{\partial(1/k_BT)} =\langle
E\rangle-\frac{\langle M^nE\rangle}{\langle M^n\rangle}.
\end{eqnarray}

Plotting these quantities as functions of $T$ for system size
($L,N_z$), we can identify the transition temperature by looking
at their respective behavior (maxima of $C_v$ and $\chi$, ...).
Note that the transition temperatures for these quantities
coincide only at infinite $L$. For large values of $L$, the
following scaling relations are expected (see details in Ref.
\onlinecite{Bunker}):

\begin{equation}
V_1^{\max}\propto L^{1/\nu}, \hspace{1cm} V_2^{\max}\propto
L^{1/\nu},
\end{equation}
\begin{equation}
C_v^{\max}=C_0+C_1L^{\alpha/\nu}\label{Cv}
\end{equation}
and
\begin{equation}
\chi^{\max}\propto L^{\gamma/\nu}
\end{equation}
at their respective 'transition' temperatures $T_c(L)$, and

\begin{equation}
C_U=C_U[T_c(\infty)]+AL^{-\alpha/\nu},
\end{equation}
\begin{equation}
M_{T_c(\infty)}\propto L^{-\beta/\nu}\label{MB}
\end{equation}
and
\begin{equation}
T_c(L)=T_c(\infty)+C_AL^{-1/\nu}\label{TC},
\end{equation}
where $A$, $C_0$, $C_1$ and $C_A$ are constants. We estimate $\nu$
independently from $V_1^{\max}$ and $V_2^{\max}$. With this value
we calculate $\gamma$ from $\chi^{\max}$ and $\alpha$ from
$C_v^{\max}$.  Note that we can estimate $T_c(\infty)$ by using
the last expression. Using $T_c(\infty)$, we can calculate $\beta$
from $M_{T_c(\infty)}$. The Rushbrooke scaling law $\alpha +2\beta
+\gamma=2$ is then in principle verified. Finally, using the
hyperscaling relationship, we can estimate the "effective"
dimension of thin films by $d_{\mbox{eff}}=(2-\alpha)/\nu$ and the
exponent $\eta$ from the scaling law $\gamma=(2-\eta)\nu$.

We note however that only $\nu$ is directly calculated from MC
data.  Exponent $\gamma$ obtained from $\chi^{\max}$ and $\nu$
suffers little errors (systematic errors and errors from $\nu$).
Other exponents are obtained by MC data and  several-step fitting.
For example, to obtain $\alpha$ we have to fit $C_v^{\max}$ of Eq.
\ref{Cv} by choosing $C_0$, $C_1$ and by using the value of $\nu$.
So in practice, in most cases, one calculates $\alpha$ or $\beta$
from MC data and uses the Rushbrooke scaling law to calculate the
remaining exponent. However, for our precise purpose we will show
in the following the results of all exponents $\nu$, $\gamma$,
$\alpha$ and $\beta$ calculated from MC data. We will show that
the Rushbrooke scaling law is very well verified. The exponent
$\alpha$ will allow us to estimate the effective dimension of the
system.

\section{Results}

We show now the results obtained by MC simulations with the
Hamiltonian (\ref{eqn:hamil1}).

\begin{figure} [h!]
\centerline{\epsfig{file=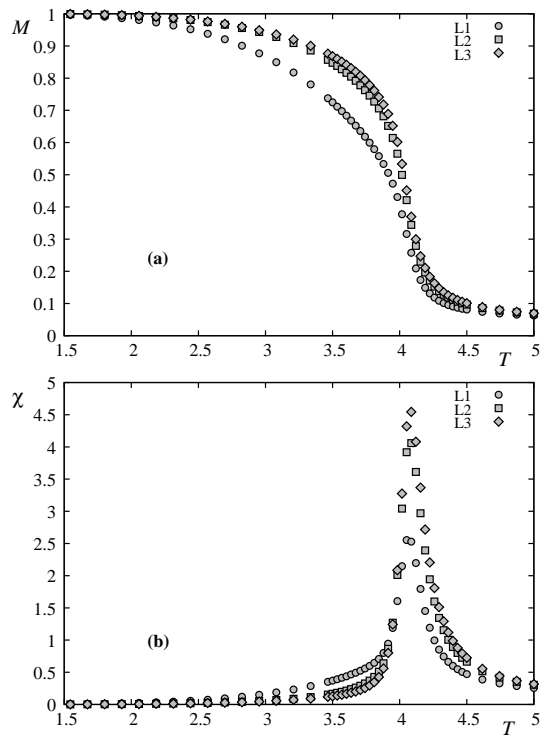,width=2.8in}} \caption{ Layer
magnetizations (a) and layer susceptibilities (b) versus $T$ with
$N_z=5$.} \label{fig:N24Z5M}
\end{figure}

Let us show in Fig. \ref{fig:N24Z5M}  the layer magnetizations and
their corresponding susceptibilities of the first three layers, in
the case where $J_s=1$.  It is interesting to note that the
surface layer is smaller that the interior layers, as it has been
shown theoretically by the Green's function method a long time
ago.\cite{diep79,diep81}  The surface spins have smaller local
field due to the lack of neighbors, so thermal fluctuations will
reduce their magnetization with respect to the interior layers.
The susceptibilities have their peaks at the same temperature,
indicating a single transition.

\begin{figure}
\centerline{\epsfig{file=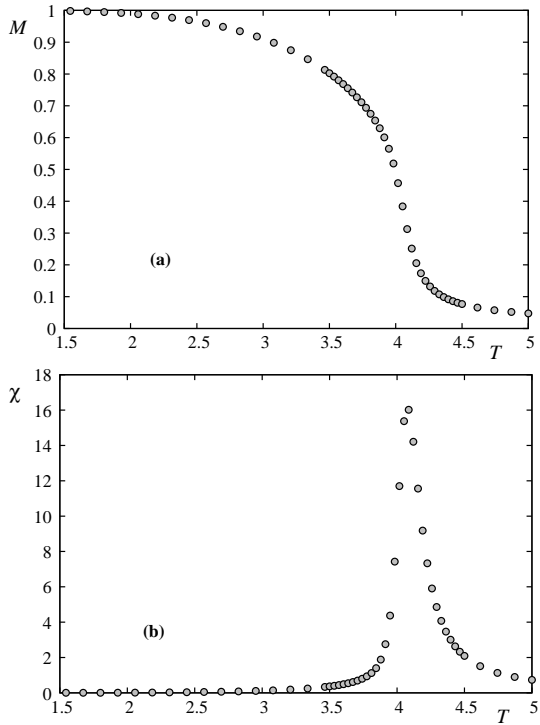,width=2.8in}} \caption{Total
magnetization (a) and total susceptibility (b) versus $T$ with
$N_z=5$.} \label{fig:N24Z5MT}
\end{figure}

Figure \ref{fig:N24Z5MT} shows total magnetization of the film and
the total susceptibility. This indicates clearly that there is
only one peak as said above.

\begin{figure}
\centerline{\epsfig{file=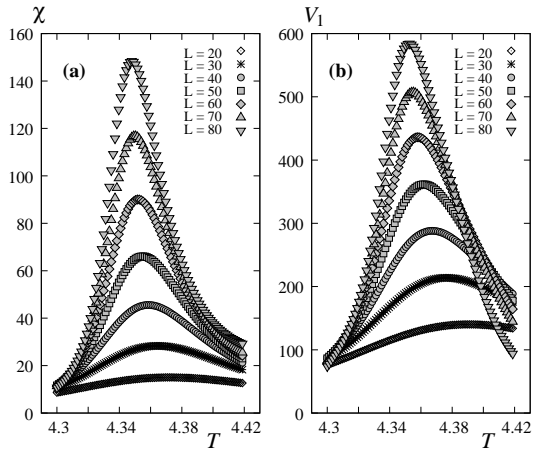,width=2.8in}}
\caption{(a) Susceptibility and (b) $V_1$, as functions of $T$ for
several $L$ with $N_z=11$, obtained by multiple histogram
technique.} \label{fig:ISSVZ11}
\end{figure}

Let us show now an example of excellent results obtained from
multiple histograms described above. Figure \ref{fig:ISSVZ11}
shows the susceptibility and the first derivative $V_1$ versus $T$
around their maxima for several sizes.

\begin{figure}
\centerline{\epsfig{file=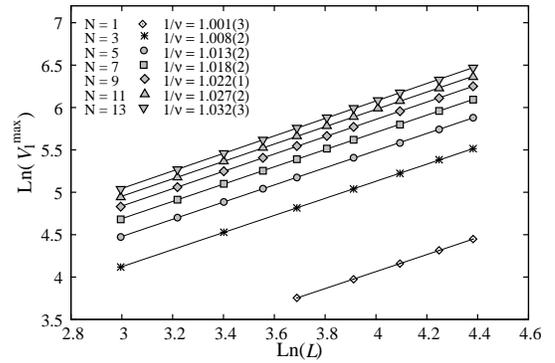,width=2.8in}} \caption{ Maximum
of the first derivative of $\ln M$  versus $ L$ in the $\ln-\ln$
scale.} \label{fig:NUL}
\end{figure}

\begin{figure}
\centerline{\epsfig{file=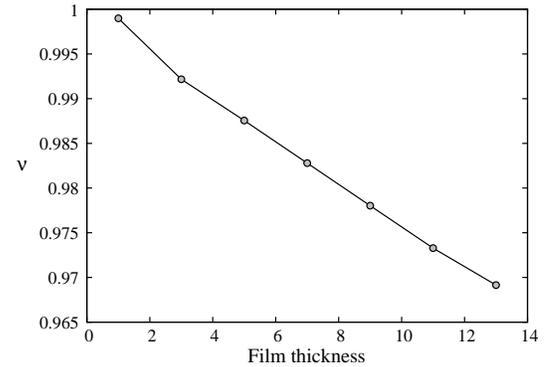,width=2.8in}} \caption{$\nu$
versus $N_z$.} \label{fig:NUZ}
\end{figure}

We show in Fig. \ref{fig:NUL} the maximum  of the first derivative
of $\ln M$ with respect to $\beta=(k_BT)^{-1}$  versus $ L$ in the
$\ln-\ln$ scale for several film thicknesses up to $N_z=13$. From
the slopes of these remarkably straight lines, we obtain $1/\nu$.
We plot in Fig. \ref{fig:NUZ} $\nu$ as a function of thickness
$N_z$. We observe here a small but systematic deviation of $\nu$
from its 2D value ($\nu_{2D}=1)$ with increasing thickness.  To
show the precision of our method, we give here the results of
$N_z=1$. For $N_z =1$, we have $1/\nu =1.0010 \pm 0.0028$ which
yields $\nu = 0.9990\pm 0.0031$ and $\gamma/\nu = 1.7537 \pm
0.0034$ (see Figs. \ref{fig:GAML} and \ref{fig:GAMZ} below)
yielding $\gamma = 1.7520\pm 0.0062$. These results are in
excellent agreement with the exact results $\nu_{2D}=1$ and
$\gamma_{2D}=1.75$.  The very high precision of our method is thus
verified in the range of the system sizes used in the present
work.

We show in Fig. \ref{fig:GAML} the maximum of the susceptibility
versus $L$ in the $\ln-\ln$ scale for  film thicknesses up to
$N_z=13$. We have used only results of $L\geq 30$. Including
$L=20$ and 25 will result, unlike the case of $\nu$,  in a
decrease of $\gamma$ of about one percent for $N_z\geq 7$.  From
the slopes of these straight lines, we obtain the values of
$\gamma/\nu$. Using the values of $\nu$ obtained above, we deduce
the values of $\gamma$ which are plotted in Fig. \ref{fig:GAMZ} as
a function of thickness $N_z$. Unlike the case of $\nu$, we
observe here a stronger deviation of $\gamma$ from its 2D value
(1.75) with increasing thickness.  This finding is somewhat
interesting: the magnitude of the deviation from the 2D value may
be different from one critical exponent to another: $\simeq 3\%$
for $\nu$ and $\simeq 8\%$ for $\gamma$ when $N_z$ goes from 1 to
13. We will see below that $\beta$ varies even more strongly.

\begin{figure}
\centerline{\epsfig{file=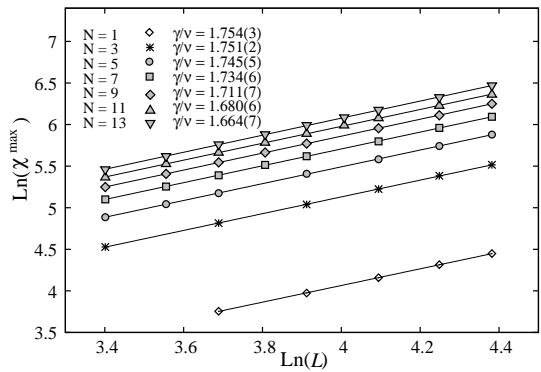,width=2.8in}} \caption{Maximum
of susceptibility versus $L$ in the $\ln-\ln$ scale.}
\label{fig:GAML}
\end{figure}

\begin{figure}
\centerline{\epsfig{file=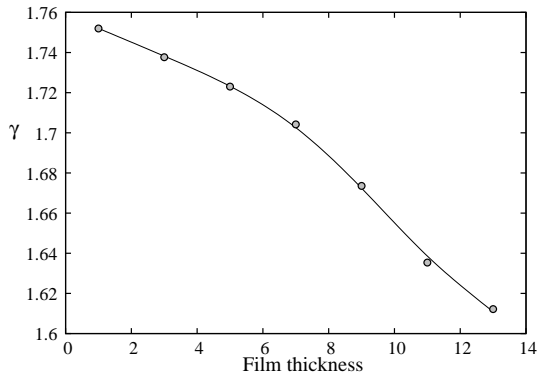,width=2.8in}}
\caption{$\gamma$ versus $N_z$.} \label{fig:GAMZ}
\end{figure}

\begin{figure}
\centerline{\epsfig{file=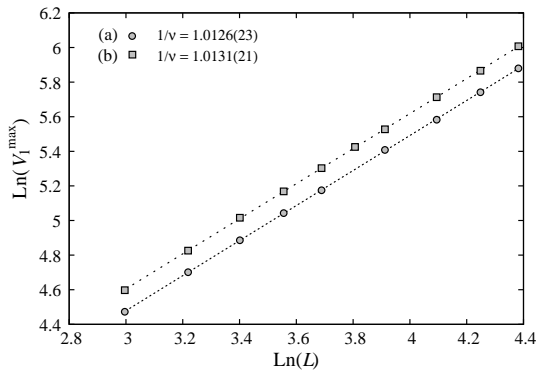,width=2.8in}} \caption{Maximum
of the first derivative of $\ln M$  versus $ L$ in the $\ln-\ln$
scale for $N_z=5$ (a) without PBC in $z$ direction (b) with PBC in
$z$ direction. } \label{fig:NUZ5}
\end{figure}

\begin{figure}
\centerline{\epsfig{file=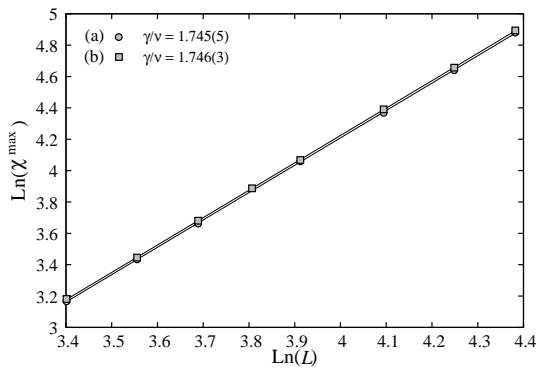,width=2.8in}}
\caption{Maximum of susceptibility versus $L$ in the $\ln-\ln$
scale for $N_z=5$ (a) without PBC in $z$ direction (b) with PBC in
$z$ direction. The points of these cases cannot be distinguished
in the figure scale.} \label{fig:GAMZ5}
\end{figure}

\begin{figure}
\centerline{\epsfig{file=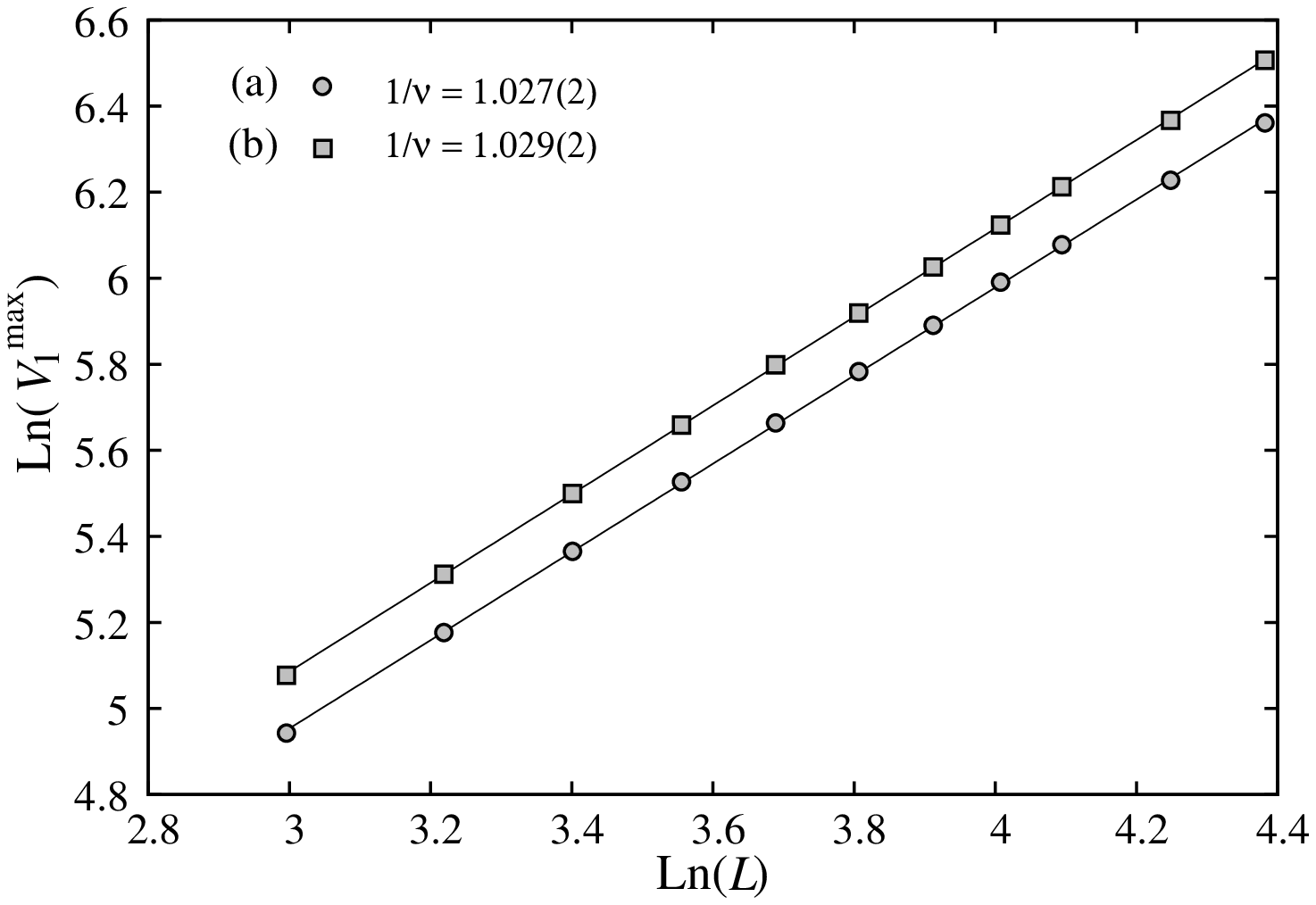,width=2.8in}}
\caption{Maximum of the first derivative of $\ln M$  versus $ L$
in the $\ln-\ln$ scale for $N_z=11$ (a) without PBC in $z$
direction (b) with PBC in $z$ direction. } \label{fig:NUZ11}
\end{figure}

\begin{figure}
\centerline{\epsfig{file=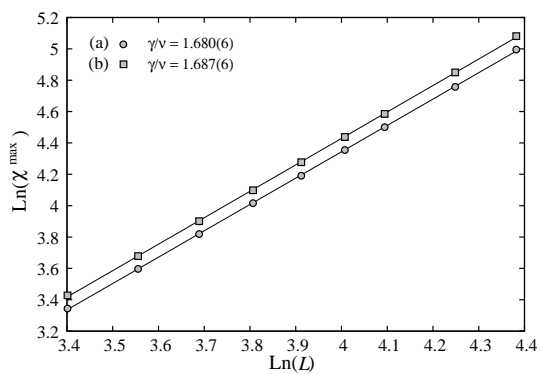,width=2.8in}}
\caption{Maximum of susceptibility versus $L$ in the $\ln-\ln$
scale for $N_z=11$ (a) without PBC in $z$ direction (b) with PBC
in $z$ direction.  } \label{fig:GAMZ11}
\end{figure}

\begin{figure}
\centerline{\epsfig{file=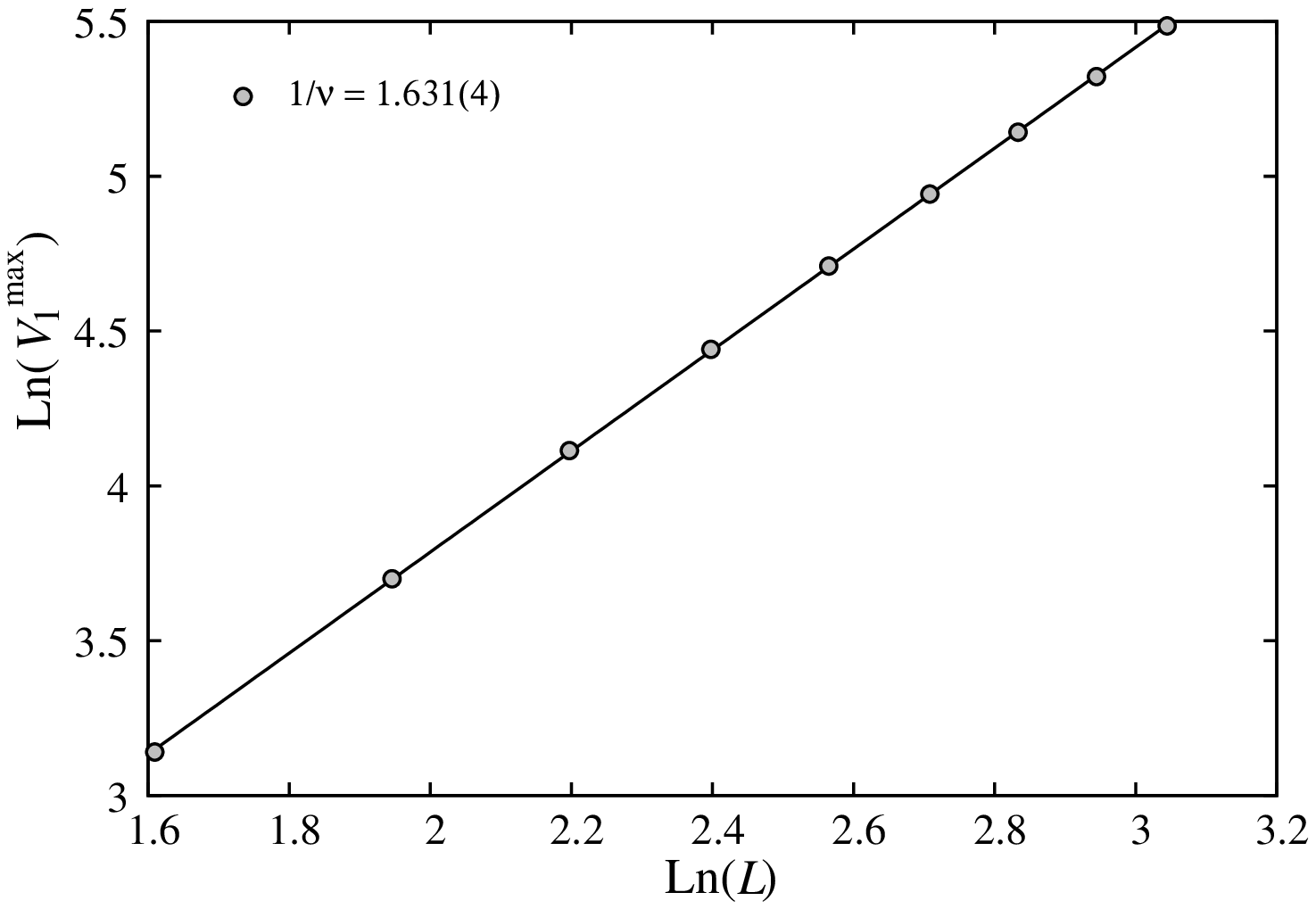,width=2.8in}} \caption{Maximum
of the first derivative of $\ln M$  versus $ L$ in the $\ln-\ln$
scale for 3D case.} \label{fig:NUL3D}
\end{figure}

\begin{figure}
\centerline{\epsfig{file=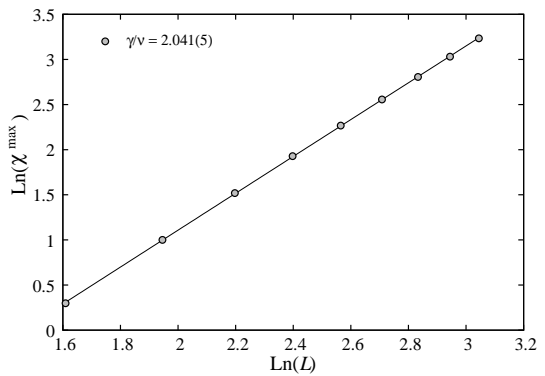,width=2.8in}}
\caption{Maximum of susceptibility versus $L$ in the $\ln-\ln$
scale for 3D case.} \label{fig:GAML3D}
\end{figure}

At this stage, a natural question arises: does the absence of PBC
in the $z$ direction cause these deviations of the critical
exponents?  The answer is no: we have calculated $\nu$ and
$\gamma$ for $N_z=5$ in both cases: with and without PBC in the
$z$ direction. The results show no significant difference between
the two cases as seen in Figs. \ref{fig:NUZ5} and \ref{fig:GAMZ5}.
We have found the same thing with $N_z=11$ shown in Figs.
\ref{fig:NUZ11} and \ref{fig:GAMZ11}. So, we conclude that the
fixed thickness will result in the deviation of the critical
exponents, not from the absence of the PBC. This is somewhat
surprising since we thought, incorrectly, that the PBC should
mimic the infinite dimension so that we should obtain the 3D
behavior when applying the PBC.  As will be seen below, the 3D
behavior is recovered only when the finite size scaling is applied
in the $z$ direction at the same time in the $xy$ plane.  To show
this, we plot in Figs. \ref{fig:NUL3D} and \ref{fig:GAML3D} the
results for the 3D case.  Even with our modest sizes (up to
$L=N_z=21$, since it is not our purpose to treat the  3D case
here), we obtain $\nu=0.613\pm0.005$ and $\gamma=1.250\pm 0.005$
very close to their 3D best known values
$\nu_{3D}=0.6289\pm0.0008$ and $\gamma_{3D}=1.2390\pm 0.0025$
obtained by using  $24\leq L \leq 96$).\cite{Ferrenberg3}

\begin{figure}
\centerline{\epsfig{file=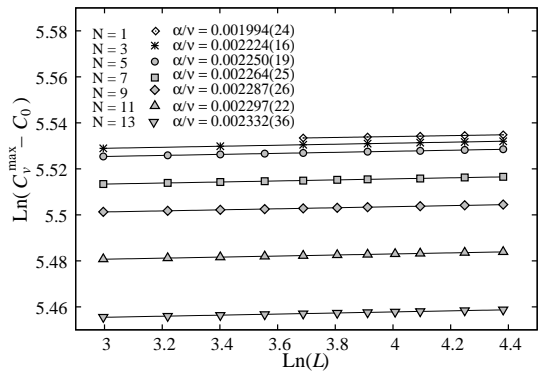,width=2.8in}} \caption{$\ln
 (C_v^{\max}-C_0)$ versus $\ln L$ for
$N_z=1,3,5,...,13$.  The slope gives $\alpha/\nu$  (see Eq.
\ref{Cv}).  } \label{fig:ALPHAL}
\end{figure}

Let us discuss on the deviation of the critical exponents due to
the film finite thickness.   For second-order transitions,
theoretical arguments, such as those
 from the
renormalization group, say that the correlation length in the
direction perpendicular  to the film is finite, hence it is
irrelevant to the criticality, the film should have the 2D
character as long as $N_z$ is finite.  We have seen above that
this is not the case here.  The deviation begins slowly as soon as
$N_z$ departs from 1.  A possible cause for the deviation  is from
the spatially non uniform correlation:  the correlation in a $xy$
plane depends obviously on its position with respect to the
surface layer.  On and near the surface, the spins suffer thermal
fluctuations more strongly than the interior spins so there is no
reason why all planes should have the same in-plane correlation
behavior even when there is no separate surface transition as in
the case $J_s=1$ studied here. Due to this spatially non uniform
fluctuations, we believe that near the phase transition, there are
simultaneously several correlation lengths which give rise to a
kind of "effective" critical exponents obtained above. Loosely
speaking, we can say in another manner that because of its spatial
non uniformity, the correlation in the direction perpendicular to
the film cannot be separately summed up, it interacts with the
$xy$ correlation giving rise to "effective" critical exponents
observed in our simulations.  In other words, the finite thickness
makes the dimension of the system something between 2 and 3.
Before showing this "effective" dimension, we show in Fig.
\ref{fig:ALPHAL} the maximum of $C_v^{\max}$ versus $L$ for
$N_z=1,3,5,...,13$.  Note that for each $N_z$ we had to look for
$C_0$, $C_1$ and $\alpha/\nu$ which give the best fit with data of
$C_v^{\max}$. Due to the fact that there are several parameters
which can induce a wrong combination of them, we impose that
$\alpha$ should satisfied the condition $0\leq \alpha \leq 0.11$
where the lower limit of $\alpha$ corresponds to the value of 2D
case and the upper limit to the 3D case.  In doing so, we get very
good results shown in Fig. \ref{fig:ALPHAL}.  From these ratios of
$\alpha/\nu$ we deduce $\alpha$ for each $N_z$. The values of
$\alpha$ are shown in Table \ref{tab:criexp} for several $N_z$.

\begin{figure}
\centerline{\epsfig{file=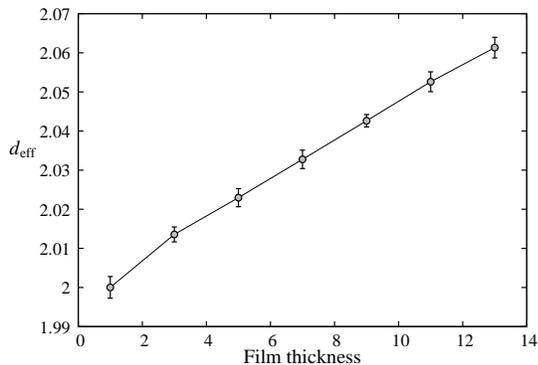,width=2.8in}}
\caption{Effective dimension of thin film as a function of
thickness. } \label{de}
\end{figure}

It is interesting to show now the effective dimension of thin film
discussed above. Replacing the values of $\alpha$ obtained above
in $d_{\mbox{eff}}=(2-\alpha)/\nu$ we obtain $d_{\mbox{eff}}$
shown in Fig. \ref{de}.

\begin{figure}
\centerline{\epsfig{file=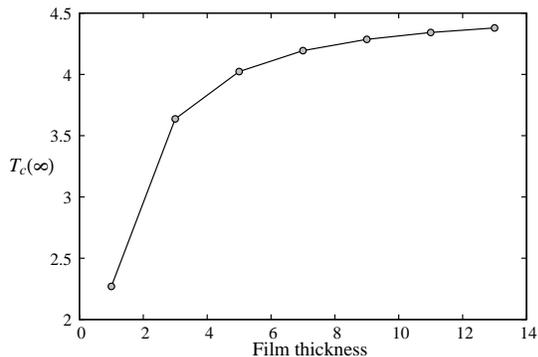,width=2.8in}}
\caption{Critical temperature at infinite $L$ as a function of the
film thickness. } \label{TCINF}
\end{figure}

We note that $d_{\mbox{eff}}$ is very close to 2. It varies from 2
to $\simeq 2.061$ for $N_z$ going from 1 to 13. The 2D character
is thus dominant even with larger $N_z$. This supports the idea
that the finite correlation in the $z$ direction, though
qualitatively causing a deviation, cannot strongly alter  the 2D
critical behavior.  This point is interesting because, as said
earlier, some thermodynamic properties may show already their 3D
values at a thickness of about a few dozens of layers, but not the
critical behavior.  To show an example of this, let us plot in
Fig. \ref{TCINF} the transition temperature at $L=\infty$ for
several $N_z$, using Eq. \ref{TC} for each given $N_z$.  As seen,
$T_c(\infty)$ reaches already $\simeq 4.379$ at $N_z=13$ while its
value at 3D is $4.51$.\cite{Ferrenberg3}  A rough extrapolation
shows that the 3D values is attained for $N_z\simeq 25$ while the
critical exponents at this thickness are far away from  the 3D
ones.

We give the precise values of $T_c(\infty)$ for each thickness.
For $N_z=1$, we have $T_c(\infty) = 2.2701 \pm 0.0003$ from $T_c$
of specific heat and  $ 2.2697 \pm 0.0005$ from $T_c$ of
susceptibility.  From these we have $T_c(\infty)=2.2699\pm
0.0005$. Note that the exact value of $T_c(\infty)$ is
 2.26919 by solving the equation $\sinh^2(2J/T_c)=1$.  Again here, the excellent
 agreement of our result shows the efficiency of the multiple histogram technique as applied
 in the present paper.
The values of $T_c(\infty)$ for other $N_z$ are summarized in
Table~\ref{tab:criexp}.

\begin{figure}
\centerline{\epsfig{file=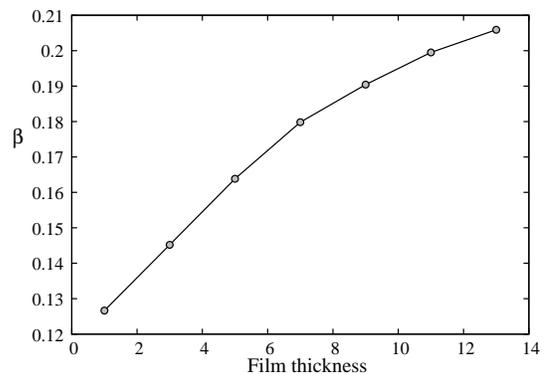,width=2.8in}}
\caption{Critical exponent $\beta$ versus the film thickness
obtained by using Eq. \ref{MB}. } \label{BETA}
\end{figure}

Calculating now $M(L)$ at these values of $T_c(\infty)$ and using
Eq. \ref{MB}, we  obtain $\beta/\nu$ for each $N_z$.  For $N_z =
1$, we have $\beta/\nu = 0.1268 \pm 0.0022$ which yields $\beta =
0.1266\pm 0.0049$ which is in excellent agreement with the exact
result (0.125). Note that if we calculate $\beta$ from $\alpha
+2\beta +\gamma =2$, then $\beta = (2-1.75198-0.00199)/2 =
0.12302\pm 0.0035$ which is in good agreement with the direct
calculation within errors.  We show in Fig. \ref{BETA} the values
of $\beta$ obtained by direct calculation using Eq.~\ref{MB}. Note
that the deviation of $\beta$ from the 2D value when $N_z$ varies
from 1 to 13 represents about 60$\%$.  Note that the 3D value of
$\beta$ is $0.3258\pm 0.0044$.\cite{Ferrenberg3}

Finally, for convenience, let us summarize our results in Table
\ref{tab:criexp} for $N_z=1,3,...,13$.  Due to the smallness of
$\alpha$, its value  is shown with 5 decimals without rounding.

\begin{widetext}
\begin{table*}
  \centering
  \caption{Critical exponents, effective dimension and critical temperature
  at infinite $xy$ limit as obtained in this paper.}\label{tab:criexp}
  \begin{tabular}{| r | c | c | c | c | c | c |}
    \hline
    $N_z$ & $\nu$ & $\gamma$ & $\alpha$ & $\beta$ & $d_{\mathrm{eff}}$ &
$T_c(\infty)$ \\
\hline 1 & $0.9990 \pm 0.0028$ & $1.7520 \pm 0.0062$ & $0.00199
\pm 0.00279$ & $0.1266 \pm 0.0049$ & $2.0000 \pm 0.0028$ &
$2.2699\pm
0.0005$ \\
3 & $0.9922 \pm 0.0019$ & $1.7377 \pm 0.0035$ & $0.00222 \pm
0.00192$ & $0.1452 \pm 0.0040$ & $2.0135 \pm 0.0019$ & $3.6365 \pm
0.0024$ \\
5 & $0.9876 \pm 0.0023$ & $1.7230 \pm 0.0069$ & $0.00222 \pm
0.00234$ & $0.1639 \pm 0.0051$ & $2.0230 \pm 0.0023$ & $4.0234 \pm
0.0028$ \\
7 & $0.9828 \pm 0.0024$ & $1.7042 \pm 0.0087$ & $0.00223 \pm
0.00238$ & $0.1798 \pm 0.0069$ & $2.0328 \pm 0.0024$ & $4.1939 \pm
0.0032$ \\
9 & $0.9780 \pm 0.0016$ & $1.6736 \pm 0.0084$ & $0.00224 \pm
0.00161$ & $0.1904 \pm 0.0071$ & $2.0426 \pm 0.0016$ & $4.2859 \pm
0.0022$ \\
11& $0.9733 \pm 0.0025$ & $1.6354 \pm 0.0083$ & $0.00224 \pm
0.00256$ & $0.1995 \pm 0.0088$ & $2.0526 \pm 0.0026$ & $4.3418 \pm
0.0032$ \\
13& $0.9692 \pm 0.0026$ & $1.6122 \pm 0.0102$ & $0.00226 \pm
0.00268$ & $0.2059 \pm 0.0092$ & $2.0613 \pm 0.0027$ & $4.3792 \pm
0.0034$ \\
    \hline
  \end{tabular}
\end{table*}
\end{widetext}

\section{Concluding remarks}

We have considered a simple system, namely the Ising model on a
simple cubic thin film,  in order to clarify the point whether or
not there is a continuous deviation of the 2D exponents with
varying film thickness.  From results obtained by the highly
accurate multiple histogram technique shown above, we conclude
that the critical exponents in thin films show a continuous
deviation from their 2D values as soon as the thickness departs
from 1. We believe that this deviation stems from deep physical
mechanisms, not from the calculation method used here. We would
like moreover to emphasize some additional interesting
observations:

1. The deviations of the exponents  from their 2D values are very
different in magnitude: while $\nu$ and $\alpha$ vary very little
over the studied range of thickness, $\gamma$ and specially
$\beta$ suffer stronger deviations

2. With a fixed thickness ($>1$),   the same critical exponents
are observed, within errors, in simulations with or without
periodic boundary condition in the $z$ direction

3. To obtain the 3D behavior, finite size scaling should be
applied simultaneously in the three directions.  If we do not
apply the scaling in the $z$ direction, we will not obtain 3D
behavior even with a very large, but fixed, thickness and even
with periodic boundary condition in the $z$ direction

4. With regard to the critical behavior, thin films behave as
systems with effective dimensions between 2 and 3, depending on
the film thickness. Note however that, except a strong deviation
of $\gamma$, other exponents stay near their 2D limit even with a
large thickness, while non critical thermodynamic properties may
attain 3D behaviors at a thickness of about a few dozens atomic
layers.

To conclude, we hope that the numerical results shown in this
paper will stimulate more theoretical analysis in search for the
origin of the continuous variation of the critical exponents with
changing thickness.  It should be also desirable to study more
cases to clarify the role of thickness on the transition behavior
of very thin films, in particular the effect of the film thickness
on the bulk first-order transition.

One of us (VTN) thanks the "Asia Pacific Center for Theoretical
Physics" (South Korea) for a financial post-doc support and
hospitality during the period 2006-2007 where part of this work
was carried out.  The authors are grateful to Yann Costes of the
University of Cergy-Pontoise for technical help in parallel
computation.

{}

\end{document}